\documentclass[prd,aps,twocolumn,superscriptaddress,dvips,10pt]{revtex4}
\usepackage{feynmp}
\usepackage{amsmath}
\usepackage{amssymb}
\usepackage{epsfig}
\usepackage{graphicx}
\usepackage{subfigure}
\usepackage{hyperref}
\usepackage{bm}
\usepackage{color}
\usepackage{lineno}
\usepackage{threeparttable}
\usepackage{supertabular}
\usepackage{booktabs,tabularx}
\begin{document}

\title{New solution to the hyperon puzzle of neutron stars: Quantum many-body effects}

\author{Hao-Fu Zhu}
\affiliation{Department of Astronomy, University of Science and Technology of China, Hefei, Anhui 230026, China}\affiliation{School of Astronomy and Space Science, University of Science and Technology
of China, Hefei, Anhui 230026, China}
\author{Guo-Zhu Liu}
\altaffiliation{Corresponding author: gzliu@ustc.edu.cn}
\affiliation{Department of Modern Physics and Anhui Center for Fundamental Sciences in Theoretical Physics, University of Science and Technology of China, Hefei, Anhui 230026, China}
\author{Xufen Wu}
\affiliation{Department of Astronomy, University of Science and
Technology of China, Hefei, Anhui 230026, China}\affiliation{School
of Astronomy and Space Science, University of Science and Technology
of China, Hefei, Anhui 230026, China}
\author{Ye-Fei Yuan}
\affiliation{Department of Astronomy, University of Science and
Technology of China, Hefei, Anhui 230026, China}\affiliation{School
of Astronomy and Space Science, University of Science and Technology
of China, Hefei, Anhui 230026, China}

\begin{abstract}
The hyperon puzzle refers to the challenge of {reconciling the
existence of hyperons} in neutron star cores and the observed high
masses of neutron stars. The recent discovery of PSR J0952-0607
($2.35\pm0.17 M_{\odot}$) has intensified this challenge. Existing
solutions fail to achieve such a high mass, and often predict
unrealistically fast cooling that is at odds with observations.
Here, we propose a novel solution to the hyperon puzzle. Using the
Dyson-Schwinger equation approach, we incorporate the quantum
many-body effects caused by strong baryon-meson interactions into
the equation of state {for cold baryonic matter} and find it stiff
enough to support a maximum hyperon star mass of {$M_{\mathrm{max}}
\approx 2.59 M_{\odot}$}, which can explain all the observed high
neutron star masses. The resulting proton and hyperon fractions are
remarkably low, thus the nucleonic and hyperonic direct Urca
processes are significantly suppressed. As a result, fast cooling
typically does not occur in ordinary neutron stars.
\end{abstract}

\maketitle

\section{Introduction \label{sec:introduction}}

Neutron stars (NSs) provide a unique platform for exploring the
intriguing behaviors of dense matter \cite{Lattimer04, Yunes22}. The
baryon density in their inner cores can be several times higher than
the nuclear saturation density $n^{}_{\mathrm{B}0}$, offering
extreme conditions unattainable in all terrestrial laboratories. In
particular, hyperons$-$baryons with strange quark content$-$have
long been conjectured to exist in NS cores due to $\beta$
equilibrium \cite{Ambartsumyan60, Glendenningbook, Glendenning82,
Glendenning92, Balberg99, Yuan05}. NSs having hyperons inside are
also referred to as hyperon stars (HSs). It is widely recognized
that NSs present the most promising environment for studying the
physical effects of hyperons.

The Walecka-type relativistic mean-field theory (RMFT)
\cite{Walecka74, Boguta77, Glendenningbook, Dutra14} has
demonstrated remarkable success in describing both finite nuclei
\cite{TM1,NL3omegarho,FSUGold,TW99,DDVT} and nuclear matter
\cite{FSUGold,BigApple,TW99,DD2,DDVT}. Consequently, it has become
the most frequently employed method of calculating the equation of
state (EOS) of NS matter, reliably reproducing a wide range of
astronomical observations \cite{Dutra16, Shen24, BigApple, Li22,
Lourenco19}. Extensive early studies based on this method have
revealed that the EOS are significantly softened after including
hyperons \cite{Glendenningbook, Glendenning82, Glendenning92,
Balberg99, Yuan05, Ban04}. This reduces the maximum NS mass
$M_{\mathrm{max}}$ down to values lower than $2.0M_{\odot}$, where
$M_{\odot}$ is the solar mass. Since 2010, several NSs with masses
exceeding $2.0M_{\odot}$ have been observed. Notable examples
include PSR J1614-2230 with a mass of $1.97\pm0.04M_{\odot}$
\cite{Demorest10}, PSR J0348+0432 with a mass of
$2.01\pm0.04M_{\odot}$ \cite{Antoniadis13}, and PSR J0740+6620 with
a mass of $2.08\pm0.07M_{\odot}$ \cite{Cromartie19}. {It is
difficult to reconcile the observations of these massive NSs with
the hypothesized existence of hyperons. This mass problem is termed
the hyperon puzzle in the NS community} \cite{Maslov15, Vidana11,
Sun23, Weissenborn12, Lonardoni15, Chatterjee16, Masuda16,
Bombaci17, Haidenbauer17, Gerstung20, Tolos20, Chorozidou24, Wei24,
Chen24, Tu25, Drago14, Drago14b, Sedrakian23, Zhang18, Fortin20,
Zdunik13, Colucci13, Li18, Long12, Li19, Miyatsu13, Lopes14,
Bombaci21, Frohaug25}.

Many possible solutions \cite{Vidana11, Weissenborn12,
Haidenbauer17, Gerstung20, Wei24, Zhang18, Fortin20, Li19,
Miyatsu13, Lopes14, Masuda16, Bombaci17, Zdunik13, Drago14,
Drago14b, Li18, Sedrakian23, Frohaug25} have been proposed to
address the hyperon puzzle. Generically, these solutions fall into
two main categories. The first category assumes the appearance of
novel degrees of freedom, such as hybrid hadron-quark phases
\cite{Masuda16, Bombaci17, Zdunik13} or $\Delta$ isobar
\cite{Drago14, Drago14b, Li18, Sedrakian23}, which can delay the
emergence of hyperons until higher densities are reached. The second
one focuses on refining the description of hyperon-involved
interactions or exploring extra repulsive interactions to stiffen
the HS EOS \cite{Vidana11, Weissenborn12, Haidenbauer17, Gerstung20,
Wei24, Zhang18, Fortin20, Li19, Miyatsu13, Lopes14, Frohaug25}. With
such manipulations, the maximum mass can be lifted to values
slightly larger than $2.0M_{\odot}$. However, despite these
advancements, the hyperon puzzle remains unsettled, as the maximum
mass computed within RMFT still falls short of the observed NS
values. {This puzzle has been further complicated by the recent
discovery of the supermassive ``black-widow'' pulsar PSR J0952-0607,
whose mass $2.35\pm0.17M_{\odot}$ \cite{Romani22} is subject to
large uncertainty because its determination relies on a number of
more complex, model-dependent astrophysical assumptions than those
for NSs in white dwarf binaries.}

In addition to the mass discrepancy mentioned above, the hyperon
puzzle usually entails a cooling inconsistency \cite{Maslov15}.
While some specific RMFT models can support NS masses in the range
of $2.2M_{\odot} - 2.3 M_{\odot}$, their consistency with
thermal-evolution observations of NSs remains an open issue
\cite{Maslov15}. Within RMFT, the symmetry energy of NS matter tends
to increase rapidly with growing baryon density. This behavior
results in a low threshold density $n_{\mathrm{nDU}}$ and,
consequently, a low threshold NS mass $M_{\mathrm{nDU}}$ above which
nucleonic direct Urca (DU) processes \cite{Lattimer91}, such as $n
\rightarrow p + e^- + \bar{\nu}_e $, are activated in a
baryon-matter core \cite{Dutra16, Fortin16, Providencia19, Lopes24,
Fortin21}. Moreover, RMFT studies predict hyperon fractions much
higher than the threshold value needed to trigger hyperonic DU
processes \cite{Prakash92}, such as $\Lambda\rightarrow
p+e+\bar{\nu}_e$. Even when the suppression from baryon pairing is
taken into account \cite{Yakovlev01, Page04}, the calculated NS
cooling still proceeds markedly faster than observed, so HSs would
become undetectable within few years, which contradicts
astrophysical observations \cite{Maslov15}.

The limitations of current theoretical approaches may originate from
the oversimplified nature of RMFT, which cannot incorporate the
impact of meson dynamics and quantum many-body effects induced by
the baryon-meson interactions. A critical investigation is needed to
examine whether including these essential features can provide
satisfactory solutions to the above two aspects of the hyperon
puzzle.

In this paper, {we demonstrate that both the mass and cooling
problems associated with the hyperon puzzle may be resolved in a
unified manner when quantum many-body effects are incorporated into
the theoretical framework.} Based on the Dyson-Schwinger (DS)
equation approach illustrated in a previous publication
\cite{Zhu24}, we carry out a quantum field-theoretical study of the
strong baryon-meson interactions. The inclusion of many-body effects
leads to sufficiently stiff EOS that permit the existence of
hyperons in the NS interior and support a maximum mass
$M_{\mathrm{max}} \approx 2.59 M_{\odot}$, which is high enough to
account for all the observed NS masses. We calculate the resulting
proton and hyperon fractions and show that they are all remarkably
low, even in the high-density regions. This prohibits nucleonic DU
processes and also substantially suppresses hyperonic DU processes.
Consequently, in our scenario, the HSs normally do not experience
rapid cooling, provided that superfluid and superconductivity are
not considered \cite{Page11, Shternin11, Zhu2410}. Our results
provide a new perspective on the internal structure of NSs and
reveal the crucial role of quantum many-body effects.

The rest of the paper is organized as follows. In
Sec.~\ref{sec:model}, we present the effective model of the HS
matter and derive the self-consistent integral equations for three
renormalization functions that account for quantum many-body
effects. In Sec.~\ref{sec:eos}, we evaluate the NS EOS and HS EOS
based on the numerical solutions of renormalization functions. In
Sec.~\ref{sec:maximummass}, we determine the maximum HS mass for
several different values of the symmetry energy slope. In
Sec.~\ref{sec:cooling}, we show the results of particle fractions
and analyze their physical influence on the fate of HS cooling rate.
A brief summary is given in Sec.~\ref{sec:summary}.

\section{Model of NS matter \label{sec:model}}

As an extension of a previous work \cite{Zhu24}, we describe the
physics of NS matter through an effective quantum hadrodynamics
model in which the baryons are coupled to three sorts of mesons
\cite{Zhu24}, including neutral $\sigma$ mesons, denoted by an
isoscalar scalar field $\sigma$, neutral vector $\omega$ mesons,
denoted by an isoscalar vector field $\omega_{\mu}= (\omega_0,
\omega_1, \omega_2, \omega_3)$, and charged vector $\rho$ mesons,
denoted by an isovector vector field $\bm{\rho}_{\mu} =
(\rho^1_{\mu}, \rho^2_{\mu}, \rho^3_{\mu})$ with
$\rho^i_{\mu}=(\rho^i_0, \rho^i_1, \rho^i_2, \rho^i_3)$. Considering
the rotational invariance around the third axis in isospin space, we
only retain the isospin three-component of $\rho^3_{\mu}$, namely,
the neutral $\rho^{0}$ mesons. Such a $\sigma$-$\omega $-$\rho$
model is represented by the following Lagrangian density:
\begin{eqnarray}
\mathcal{L} = \mathcal{L}_{\mathrm{Baryon}} +
\mathcal{L}_{\mathrm{meson}}+ \mathcal{L}_{\mathrm{lepton}},
\label{eq:totallagrangian}
\end{eqnarray}
where
\begin{eqnarray}
\mathcal{L}_{\mathrm{Baryon}} &=&
\sum_\mathrm{B}\overline{\psi}_\mathrm{B}
\big(i\partial_{\mu}\gamma^{\mu}-m_\mathrm{B}+g_{\sigma
\mathrm{B}}\sigma - g_{\omega\mathrm{B}}
\omega_{\mu}\gamma^{\mu} \nonumber \\
&& -\Gamma_{\rho B}\rho^3_{\mu}I_{3\mathrm{B}}
\gamma^{\mu}\big)\psi^{}_\mathrm{B}, \nonumber \\
\mathcal{L}_{\mathrm{meson}} &=& \frac{1}{2}\partial_{\mu}
\sigma\partial^{\mu}\sigma - \frac{1}{2}m^{\ast2}_\sigma\sigma^2
-\frac{1}{4}\omega_{\mu\nu}\omega^{\mu\nu} \nonumber \\
&& +\frac{1}{2} m^{2}_{\omega}\omega_{\mu}\omega^{\mu}-
\frac{1}{4}\rho^3_{\mu\nu}\rho^{3\mu\nu} +
\frac{1}{2}m^{2}_{\rho} \rho^3_{\mu}\rho^{3\mu} ,
\nonumber \\
\mathcal{L}_{\mathrm{lepton}} &=& \sum_{l}
\overline{\psi}_l\left(i\partial_{\mu} \gamma^{\mu} -
m_l\right)\psi^{}_{l}, \label{eq:Lhyperonmodel}
\end{eqnarray}
where $\partial_{\mu} = (\partial_{t},\bm{\partial})$, $\gamma^{\mu}
= (\gamma^{0},\bm{\gamma})$, $\omega_{\mu\nu} =
\partial_{\mu}\omega_{\nu} -
\partial_{\nu}\omega_{\mu}$, and $\rho^3_{\mu\nu} = \partial_{\mu}
\rho^3_{\nu}-\partial_{\nu}\rho^3_{\mu}$. The spinor
$\psi^{}_\mathrm{B}$, whose conjugate is
$\overline{\psi}_\mathrm{B}=\psi^{\dagger}_\mathrm{B}\gamma^{0}$,
has four components for baryonic matter, and {the summation on
$\mathrm{B}$ is over all the charge states of the baryon octet}
$\mathrm{B} = (n, p,\Lambda,\Sigma^{+}, \Sigma^{-}, \Sigma^{0},
\Xi^{-}, \Xi^{0})$. The baryons couple to $\sigma$, $\omega$, and
$\rho^0$ mesons. The isospin projection in isospin space is denoted
by $I_{3\mathrm{B}} =
\mathrm{diag}\left(-\frac{1}{2},\frac{1}{2},0,1,-1,0,-\frac{1}{2},
\frac{1}{2}\right)$, which is the matrix containing the isospin
charges. Bare baryon masses are $m_{\mathrm{n,p}} =939~$,
$m_\Lambda=1116~$, $m_{\Sigma^{0,\pm}}=1193~$MeV, and
$m_{\Xi^{0,-}}=1318~$MeV. Rest meson masses are $m_{\sigma} = 550~$,
$m_{\omega} = 783~$, and $m_{\rho} = 763~$MeV. $m^{\ast}_{\sigma}$
denotes the renormalized mass of $\sigma$. Leptons $\psi^{}_{l}$,
where $l=e^{-},\mu^{-}$, are included to ensure the $\beta$
equilibrium and electrical neutrality. The rest lepton masses are
$m_{e} = 0.511~$ and $m_{\mu} = 105.7~$MeV.

The pronounced isospin asymmetry of NSs renders their properties,
such as the crust structure, radius, tidal deformability, and the
thresholds for DU processes, highly sensitive to the symmetry energy
slope $L_{\mathrm{s}}$ \cite{Oyamatsu07, Cavagnoli11, Grill12,
Lopes24, Shen21, Shen19, Zhu24}, which is strongly influenced by
isovector mesons. To systematically investigate how $L_{\mathrm{s}}$
affects properties of NSs, we introduce the density-dependent
isovector coupling parameter $\Gamma_{\rho \mathrm{B}}$, which has
been widely adopted in density-dependent RMFT models
\cite{TW99,DD2,DDVT,Shen21} and takes the form
\begin{eqnarray}
\Gamma_{\rho \mathrm{B}} = g_{\rho \mathrm{B}} \exp\left[-a_{\rho}
\left(\frac{n^{\ast}_{\mathrm{F}}}{ n^{~}_{\mathrm{B0}}} -
1\right)\right].
\end{eqnarray}
Here, $g_{\rho\mathrm{B}}$ is a density-independent coupling
constant, $n^{\ast}_{\mathrm{F}}$ is the total baryon density, and
$a_{\rho}$ is a tuning parameter. At the nuclear saturation density
$n^{\ast}_{\mathrm{F}} = n^{}_{\mathrm{B}0}$, $\Gamma_{\rho
\mathrm{B}}$ reverts back to $g_{\rho\mathrm{B}}$. Notably, the
effective coupling of the $\rho^0$ meson to baryons is governed by
the product $g_{\rho B} I_{3B}$. Previous studies \cite{Zhu24,
Shen21, Shen19} have revealed that increasing $a_{\rho}$ markedly
lowers the symmetry energy slope $L_{\mathrm{s}}$. A smaller
$L_{\mathrm{s}}$ would soften the EOS of the NS core, which yields
more compact NSs in the intermediate-mass regime and leaves the
maximum mass nearly unchanged. The other two nucleon-meson coupling
constants $g_{\sigma\mathrm{B}}$ and $g_{\omega\mathrm{B}}$ are
density independent.

\begin{figure}[htbp]
\centering
\includegraphics[width=3.3in]{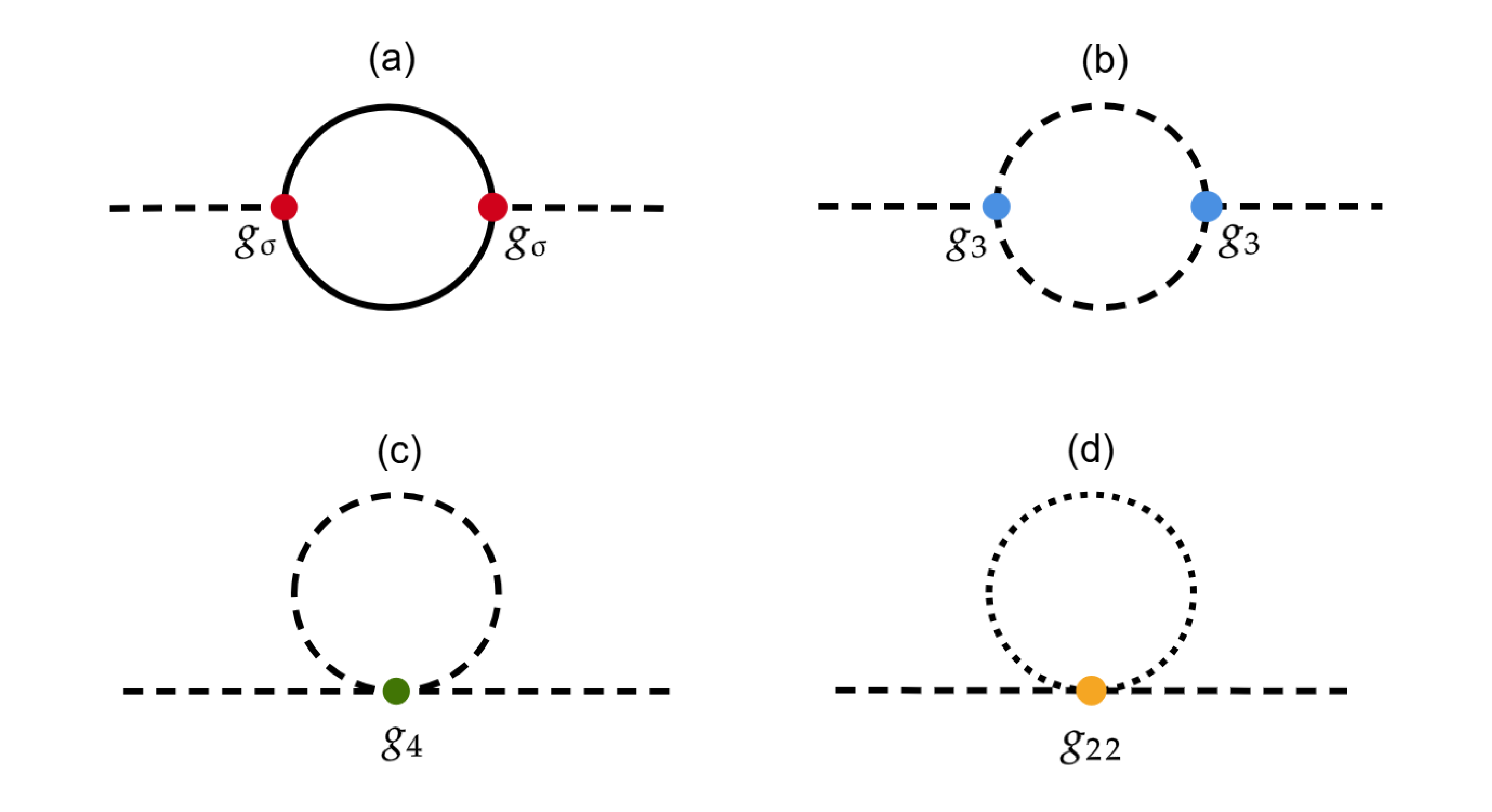}
\caption{Feynman diagrams for one-loop self-energy corrections to
the $\sigma$ meson mass. Solid line represents free baryon
propagator. Dashed (dotted) line represents free $\sigma$ ($\omega$)
meson propagator. Corrections of (a)-(d) come from baryon-$\sigma$
coupling, self-coupling $g_{3}\sigma^{3}$, self-coupling
$g_{4}\sigma^{4}$, and cross-coupling $g_{22}\sigma^{2}\omega^{2}$,
respectively.} \label{FeymannDiagram}
\end{figure}

In addition to baryon-meson couplings, mesons can couple to
themselves and to each other through terms such as
$-g_{3}\sigma^{3}$, $-g_{4}\sigma^{4}$, and
$-g_{22}\sigma^{2}\omega^{2}$, etc. These nonlinear self- and
cross-couplings appear in almost all RMFT models and play a crucial
role in the determination of a realistic EOS. However, their
coefficients are frequently negative, which can drive the
thermodynamic potential unbounded from below and trigger an
instability \cite{Glendenningbook, Zhu24}. We wish to retain the
contributions of such nonlinear couplings while eliminating any risk
of instability. According to the generic principles of quantum field
theory, the main effect of meson self- and cross-couplings is the
renormalization of bare meson masses. To illustrate this, consider
the $\sigma$ meson as an example. At one-loop level, its self-energy
receives contributions from the baryon-$\sigma$ couplings and the
nonlinear meson couplings, with the corresponding diagrams shown in
Fig.~\ref{FeymannDiagram}. These corrections shift the bare $\sigma$
mass $m_{\sigma}$ to an effective renormalized mass
$m_{\sigma}^{\ast}$, whose value depends on $g_{\sigma}$, $g_{3}$,
$g_{4}$, and $g_{22}$. This implies that the influence of $\sigma$
self- and cross-couplings between $\sigma$ and $\omega$ mesons is
packaged into a single quantity $m_{\sigma}^{\ast}$. Guided by this
consideration, we replace the bare mass $m_{\sigma}$ in
$\mathcal{L}_{\mathrm{meson}}$ by the renormalized mass
$m^{\ast}_{\sigma}$ from the outset and treat the ratio
$m^{\ast}_{\sigma}/m_{\sigma}$ as an adjusting parameter, whose
value will be determined by fitting to nuclear saturation properties
\cite{Zhu24}. This substitution eliminates potential instability
while packaging the physical effects of nonlinear meson interactions
compactly into $m_{\sigma}^{\ast}$. At the same time,
$m_{\sigma}^{\ast}$ automatically incorporates the feedback of
baryon-$\sigma$ coupling through the processes given by
Fig.~\ref{FeymannDiagram}(a). The $\omega$ and $\rho^0$ masses can
be modified by similar nonlinear couplings, but the modifications
are numerically negligible since their bare masses ($m_{\omega}$ and
$m_{\rho}$) are already quite large. We therefore fix $m_{\omega}$
and $m_{\rho}$ at their bare values.

Throughout the following calculations, NSs are treated as
approximately at zero temperature, as the typical baryon Fermi
energies ($E_{\mathrm{FB}}\approx$ 100-1000 $\mathrm{MeV}$) in NSs
far exceed the characteristic thermal energies ($k_{B}T \approx$
0.01-0.1 $\mathrm{MeV}$). We anticipate that strong baryon-meson
interactions will lead to significant quantum many-body effects,
such as the Landau damping of baryons, the baryon velocity
renormalization, and the baryon mass renormalization. To incorporate
these effects into the EOS, we will handle the model of
Eq.~(\ref{eq:Lhyperonmodel}) by employing the field-theoretical
approach developed in \cite{Zhu24}. The essence of this approach is
to calculate the EOS based on the solutions of the DS equation of
renormalized baryon propagator $G_\mathrm{B}(k)$. As shown in
\cite{Zhu24}, $G_\mathrm{B}(k)$ satisfies the following DS equation
\begin{eqnarray}
G^{-1}_\mathrm{B}(k) &=& G_{\mathrm{B0}}^{-1}(k)-ig^2_{\sigma
\mathrm{B}}\int\frac{d^4q}{(2\pi)^4}G_\mathrm{B}(k+q)D_0(q)
\nonumber \\
&&-ig^2_{\omega \mathrm{B}}\gamma_\mu\int\frac{d^4q}{(2\pi)^4}
G_\mathrm{B}(k+q)F^{\mu\nu}_{0}(q)\gamma_{\nu} \nonumber \\
&&-i\Gamma^2_{\rho \mathrm{B}}I_{3\mathrm{B}}\gamma_\mu\int\frac{d^4q}{(2\pi)^4}
G_\mathrm{B}(k+q)V^{\mu\nu}_{0}(q)I_{3\mathrm{B}}\gamma_{\nu}.\nonumber \\
\label{eq:finalme}
\end{eqnarray}
The free baryon propagator is
\begin{eqnarray}
G_{\mathrm{B}0}(k) =
\frac{1}{k_{\mu}\gamma^{\mu}-m^{}_{\mathrm{B}}},
\end{eqnarray}
and three free propagators of $\sigma$, $\omega$, and $\rho^0$
mesons, which are listed in order as follow:
\begin{eqnarray}
D_0(q) &=& \frac{1}{q^{2}-m^{\ast2}_{\sigma}}, \\
F^{\mu\nu}_0(q) &\approx& -\frac{g^{\mu\nu}}{{q}^{2}-m^{2}_{\omega}}, \\
V^{\mu\nu}_0(q) &\approx& -\frac{g^{\mu\nu}}{{q}^{2}-m^{2}_{\rho}}.
\label{eq:dselwmodelhyperon}
\end{eqnarray}
The four-momenta of neutrons and mesons are
$k\equiv(\varepsilon,\mathbf{k})$ and $q\equiv(\omega,\mathbf{q})$,
respectively. The meson propagators are functions of both the energy
and momentum, thus the DS equation (\ref{eq:finalme}) incorporates
the dynamics of all mesons, which is neglected in mean-field
theories.  Moreover, we have dropped the terms
$q^{\mu}q^{\nu}/m^{2}_{\omega}$ from $F_{0}^{\mu\nu}(q)$ and
$q^{\mu}q^{\nu}/m^{2}_{\rho}$ from $V_{0}^{\mu\nu}(q)$ to preserve
the baryon number conservation and isospin conservation,
respectively. On account of the translational invariance and
the rotational symmetry of infinite baryonic matter in the rest frame, we retain solely the time component of $\omega_\mu$ and $\rho^3_{\mu}$, which is achieved by
\begin{eqnarray}
\gamma_{\mu} \rightarrow \gamma_0, \quad \quad \quad
I_{3\mathrm{B}}\gamma_{\mu} \rightarrow I_{3\mathrm{B}}\gamma_0, \\
F_0^{\mu\nu}(q)\approx F_{0}^{00}(q), \quad
V_{0}^{\mu\nu}(q) \approx V_0^{00}(q).
\end{eqnarray}

The Fermi energy $E^{~}_{\mathrm{FB}}$ of baryons provides a natural energy scale \cite{Zhu24} and will be used to define the integration
range of $\omega$. Here, we choose $\omega \in
[-\Omega_{\mathrm{c}},+\Omega_{\mathrm{c}}]$, where
$\Omega_{\mathrm{c}}=1000~\mathrm{MeV}$ is of the same order of
$E^{~}_{\mathrm{FB}}$ at $6n^{~}_{\mathrm{B}0}$.
{We have verified through numerical calculations that the EOS and NS properties are virtually insensitive to the precise value of $\Omega_{\mathrm{c}}$ for $\Omega_{\mathrm{c}} \ge 1000~\mathrm{MeV}$.} The absolute value of the meson momentum $|\mathbf{q}|$ lies within the range of $[0,\Lambda_{\mathrm{c}}k^{~}_{\mathrm{FB}}]$, where $k^{~}_{\mathrm{FB}}$ is the Fermi momentum of baryons and $\Lambda_{\mathrm{c}}$ is a positive tuning parameter \cite{Zhu24}. {The parameter $\Lambda_{\mathrm{c}}$ is constrained by fitting the saturation properties of nuclear matter.} The integral measure is expressed as
\begin{eqnarray}
\int \frac{d^4q}{(2\pi)^4} \equiv
\int_{-\Omega_{\mathrm{c}}}^{+\Omega_{\mathrm{c}}}
\frac{d\omega}{2\pi}\int_{0}^{\Lambda_{\mathrm{c}}
k^{~}_{\mathrm{FB}}} \frac{d^{3}\mathbf{q}}{(2\pi)^{3}}.
\end{eqnarray}
All the results are free of divergences, thus renormalization calculations are not needed.

{In infinite baryonic matter, we further assume invariance under parity and time reversal, in addition to the translational and rotational symmetries already imposed above. Under these symmetries, tensor or pseudoscalar terms are forbidden in the baryon propagator.} We define three functions $A_{0\mathrm{B}}(k)$, $A_{1\mathrm{B}}(k)$, and $A_{2\mathrm{B}}(k)$ to manifest the Landau damping, velocity renormalization, and mass renormalization, respectively. Then the baryon propagator $G_\mathrm{B}(k)$ can be expressed in a generic form
\begin{eqnarray}
G_\mathrm{B}(k) = \frac{1}{A_{0\mathrm{B}}(k) \varepsilon \gamma^0 -
A_{1\mathrm{B}}(k)\mathbf{k} \cdot\bm{\gamma} -
A_{2\mathrm{B}}(k)m^{~}_{\mathrm{B}}}.
\label{eq:genericformgphyperon}
\end{eqnarray}
$A_{0\mathrm{B}}(k)$, $A_{1\mathrm{B}}(k)$, and $A_{2\mathrm{B}}(k)$ are equal to unity in the noninteracting limit, but driven by baryon-meson interactions to deviate from unity. According to our numerical calculations \cite{Zhu24}, $A_{0\mathrm{B}}(k)$, $A_{1\mathrm{B}}(k)$, and $A_{2\mathrm{B}}(k)$ exhibit a rather weak dependence on the momentum $\mathbf{k}$ for a fixed energy. It is therefore justified \cite{Zhu24} to fix their $|\mathbf{k}|$ at the Fermi momentum $k^{~}_{\mathrm{F}}$. Then,
$A_{0\mathrm{B}}(\varepsilon)$, $A_{1\mathrm{B}}(\varepsilon)$, and $A_{2\mathrm{B}}(\varepsilon)$ depend solely on the energy $\varepsilon$. Substituting Eq.~(\ref{eq:genericformgphyperon}) into the DS equation (\ref{eq:finalme}) leads to three self-consistent integral equations,
\begin{widetext}
\begin{eqnarray}
A_{0\mathrm{B}}(\varepsilon) &=& 1-\frac{i}{\varepsilon}\int\frac{d\omega
d^3\mathbf{q}}{(2\pi)^4}\frac{A_{0\mathrm{B}}(\varepsilon+\omega)(\varepsilon
+\omega)}{A^{2}_{0\mathrm{B}}(\varepsilon+\omega)(\varepsilon+\omega)^{2}
- A^{2}_{1\mathrm{B}}(\varepsilon+\omega)(\mathbf{k}+\mathbf{q})^{2}
- A^{2}_{2\mathrm{B}}(\varepsilon+\omega)m^{2}_\mathrm{B}} \nonumber \\
&&\times\Big(\frac{g^{2}_{\sigma\mathrm{B}}}{\omega^{2}-\mathbf{q}^{2}
- m^{\ast2}_{\sigma}} - \frac{g^{2}_{\omega\mathrm{B}}}{\omega^{2} -
\mathbf{q}^{2}-m^{2}_{\omega}} - \frac{\Gamma^{2}_{\rho\mathrm{B}}
I^2_{3\mathrm{B}}}{\omega^{2}-\mathbf{q}^{2}-m^{2}_{\rho}}\Big),
\label{eq:integraleqA0hyperon} \\
A_{1\mathrm{B}}(\varepsilon) &=& 1-\frac{i}{|\mathbf{k}|}\int
\frac{d\omega d^3\mathbf{q}}{(2\pi)^4}
\frac{A_{1\mathrm{B}}(\varepsilon+\omega)|\mathbf{k} +
\mathbf{q}|}{A^{2}_{0\mathrm{B}}(\varepsilon+\omega)(\varepsilon +
\omega)^{2} - A^{2}_{1\mathrm{B}}(\varepsilon+\omega)
(\mathbf{k}+\mathbf{q})^{2} - A^{2}_{2\mathrm{B}}(\varepsilon +
\omega)m^{2}_\mathrm{B}} \nonumber \\
&&\times\Big(\frac{g^{2}_{\sigma\mathrm{B}}}{\omega^{2}-\mathbf{q}^{2}
- m^{\ast2}_{\sigma}} + \frac{g^{2}_{\omega\mathrm{B}}}{\omega^{2} -
\mathbf{q}^{2}-m^{2}_{\omega}} + \frac{\Gamma^{2}_{\rho\mathrm{B}}
I^2_{3\mathrm{B}}}{\omega^{2}-\mathbf{q}^{2}-m^{2}_{\rho}}\Big),
\label{eq:integraleqA1hyperon}\\
A_{2\mathrm{B}}(\varepsilon)&=& 1+\frac{i}{m_{\mathrm{B}}}\int
\frac{d\omega d^3\mathbf{q}}{(2\pi)^4}
\frac{A_{2\mathrm{B}}(\varepsilon+\omega) m_{\mathrm{B}}}{A^{2}_{0
\mathrm{B}}(\varepsilon+\omega)(\varepsilon+\omega)^{2} -
A^{2}_{1\mathrm{B}}(\varepsilon+\omega)(\mathbf{k}+\mathbf{q})^{2} -
A^{2}_{2\mathrm{B}}(\varepsilon+\omega)m^{2}_\mathrm{B}} \nonumber \\
&&\times\Big(\frac{g^{2}_{\sigma\mathrm{B}}}{\omega^{2}-\mathbf{q}^{2}
- m^{\ast2}_{\sigma}}-\frac{g^{2}_{\omega\mathrm{B}}}{\omega^{2} -
\mathbf{q}^{2}-m^{2}_{\omega}}-\frac{\Gamma^{2}_{\rho\mathrm{B}}
I^2_{3\mathrm{B}}}{\omega^{2}-\mathbf{q}^{2}-m^{2}_{\rho}}\Big).
\label{eq:integraleqA2hyperon}
\end{eqnarray}
\end{widetext}
The three functions $A_{0\mathrm{B}}(\varepsilon)$, $A_{1\mathrm{B}}(\varepsilon)$, and $A_{2\mathrm{B}}(\varepsilon)$ can be determined by numerically solving the above three equations using the iteration method \cite{Zhu24}. {It is challenging to compute the EOS directly from these energy-dependent functions. To simplify the calculation,} we find it convenient to average over the energies of $A_{0\mathrm{B}}(\varepsilon)$, $A_{1\mathrm{B}}(\varepsilon)$, and $A_{2\mathrm{B}}(\varepsilon)$. The average is carried out as follows
\begin{eqnarray}
{\bar A}_{0\mathrm{B},1\mathrm{B},2\mathrm{B}} = \frac{\int A_{0\mathrm{B},1\mathrm{B},2\mathrm{B}}(\varepsilon) d\varepsilon}{\int d\varepsilon}.
\end{eqnarray}
These three quantities depend on the baryon density and take into account the quantum many-body effects caused by strong baryon-meson interactions. Then the original Lagrangian density for the baryon sector, namely $\mathcal{L}_{\mathrm{Baryon}}$, is renormalized to become
\begin{eqnarray}
\widetilde{\mathcal{L}}_{\mathrm{Baryon}} &=& \sum_{\mathrm{B}}
\overline{\psi}_\mathrm{B}\big(i\overline{A}_{0\mathrm{B}}
\partial_{t}\gamma^{0} + i\overline{A}_{1\mathrm{B}}\bm{\partial}\cdot
\bm{\gamma} - \overline{A}_{2\mathrm{B}}m_{\mathrm{B}} \nonumber
\\
&&+g_{\sigma \mathrm{B}}\sigma-g_{\omega \mathrm{B}}\omega_{\mu}
\gamma^{\mu}-\Gamma_{\rho B}\rho^{3}_{\mu} I_{3\mathrm{B}}
\gamma^{\mu}\big)\psi^{}_{\mathrm{B}}.
\label{eq:renormalizedLhyperon}
\end{eqnarray}

The hyperon-meson coupling parameters are treated following the
usual approach \cite{Chiapparini09}. Define several ratios,
\begin{eqnarray}
x_{\sigma \mathrm{B}} = \frac{g_{\sigma \mathrm{B}}}{g_{\sigma
\mathrm{N}}}, \quad x_{\omega \mathrm{B}} = \frac{g_{\omega
\mathrm{B}}}{g_{\omega \mathrm{N}}}, \quad x_{\rho \mathrm{B}} =
\frac{g_{\rho \mathrm{B}}}{g_{\rho \mathrm{N}}}.
\end{eqnarray}
For the couplings with $\omega$ and $\rho^0$ mesons, we utilize the
SU(6) symmetry relations \cite{Dover84}:
\begin{eqnarray}
&& x_{\omega\Lambda} = x_{\omega\Sigma}=\frac{2}{3}, \quad x_{\omega
\Xi}=\frac{1}{3},\label{relationomega} \\
&& x_{\rho\Lambda} = x_{\rho\Sigma}= x_{\rho\Xi}=1.
\label{relationrho}
\end{eqnarray}
The coupling constants for hyperon-$\sigma$ meson can be obtained
from hypernuclear potentials:
\begin{eqnarray}
V_\Lambda &=& x_{\omega\Lambda}V_{\omega
\mathrm{N}}-x_{\sigma\Lambda}V_{\sigma \mathrm{N}}= -28~
\mathrm{MeV},\label{relationsigma1} \\
V_\Sigma &=& x_{\omega\Sigma} V_{\omega \mathrm{N}}-x_{\sigma\Sigma}
V_{\sigma \mathrm{N}}= +30~ \mathrm{MeV},\label{relationsigma2} \\
V_\Xi &=& x_{\omega\Xi} V_{\omega \mathrm{N}}-x_{\sigma\Xi}
V_{\sigma \mathrm{N}}= -18~ \mathrm{MeV},\label{relationsigma3}
\end{eqnarray}
where $V_{\omega \mathrm{N}}=g_{\omega \mathrm{N}}\omega_0$ and
$V_{\sigma \mathrm{N}}=g_{\sigma \mathrm{N}}\sigma$ are the nuclear
potentials for saturated symmetric nuclear matter \cite{Bielich02,
Friedman07}. Combining Eqs.~(\ref{relationomega})$-$(\ref{relationsigma3}),
we can ultimately obtain
\begin{eqnarray}
x_{\sigma \Lambda } =0.5969,~ x_{\sigma\Sigma} =0.4223,~
x_{\sigma\Xi } =0.3105.
\end{eqnarray}

\section{Equation of state \label{sec:eos}}

In this section, we calculate the EOS of the NS matter. For this
purpose, we replace $\mathcal{L}_{\mathrm{Baryon}}$ appearing in
Eq.~(\ref{eq:totallagrangian}) with
$\widetilde{\mathcal{L}}_{\mathrm{Baryon}}$ given by
Eq.~(\ref{eq:renormalizedLhyperon}) and then compute the
energy density and pressure by adopting the standard procedure of
RMFT \cite{Glendenningbook}. The quantum many-body effects resulting
from the baryon-meson interactions are already incorporated in the
three averaged quantities $\overline{A}_{0\mathrm{B}}$,
$\overline{A}_{1\mathrm{B}}$, and $\overline{A}_{2\mathrm{B}}$.

The equation of motion of the baryon field has the form
\begin{eqnarray}
&& \left[i\bar{A}_{0\mathrm{B}}\partial_t\gamma^0 +
i\bar{A}_{1\mathrm{B}}\bm{\partial} \cdot \bm{\gamma} -
\bar{A}_{2\mathrm{B}} m_{\mathrm{B}}^{}\right]\psi_\mathrm{B}(z)
\nonumber \\
&=& -g_{\sigma\mathrm{B}} \sigma(z)\psi_{\mathrm{B}}(z)+ g_{\omega\mathrm{B}}
\omega_{\mu}(z) \gamma^{\mu}\psi_\mathrm{B}(z) \nonumber \\
&& +\Gamma_{\rho\mathrm{B}}\rho^{3}_{\mu}(z)
I_{3\mathrm{B}}\gamma^{\mu}\psi_\mathrm{B}(z) +
\Sigma^\mathrm{R}_{\mu}\gamma^{\mu}\psi_\mathrm{B}(z).
\label{eq:eomnucleon}
\end{eqnarray}
where the density dependence of $\Gamma_{\rho\mathrm{B}}$
contributes a rearrangement term \cite{Lenske95} for baryons:
\begin{eqnarray}
\Sigma^\mathrm{R}_{\mu} &=& \frac{J_\mu}{n^{\ast}_{\mathrm{F}}}
\sum_\mathrm{B}\frac{\partial\Gamma_{\rho\mathrm{B}}}{\partial
n^{\ast}_{\mathrm{F}}}\rho^{3}_{\nu}\overline{\psi}_\mathrm{B}
 I_{3\mathrm{B}}\gamma^{\nu}\psi_\mathrm{B}.
\end{eqnarray}
{Here, the total baryon current is given by
\begin{eqnarray}
J_{\mu} &=& \sum_{\mathrm{B}} \bar{\psi}_{\mathrm{B}}
\gamma_{\mu}\psi_\mathrm{B},
\end{eqnarray}
and the total baryon density is
\begin{eqnarray}
n^{\ast}_\mathrm{F} = \sum_\mathrm{B}n^{\ast}_{\mathrm{B}},
\end{eqnarray}
in which the renormalized baryon density of each species is
\begin{eqnarray}
n^{\ast}_\mathrm{B} = 2\int^{k_{\mathrm{FB}}}_{0}
\frac{d^3\mathbf{k}}{(2\pi)^{3}}\frac{1}{\bar{A}_{0\mathrm{B}}}.
\end{eqnarray}}

The equations of motion of the three meson fields are of the forms
\begin{eqnarray}
\left(\partial_{\mu}\partial^{\mu}+m^{\ast2}_{\sigma}\right)\sigma(z)
&=& \sum_\mathrm{B}g_{\sigma\mathrm{B}}\bar{\psi}_{\mathrm{B}}(z)
\psi_\mathrm{B}(z), \label{eq:sigmaeom} \\
\partial_{\mu}\omega^{\mu\nu}(z)+m^{2}_{\omega}\omega^{\nu}(z) &=&
\sum_\mathrm{B}g_{\omega\mathrm{B}}\bar{\psi}_\mathrm{B}(z)
\gamma^{\nu}\psi_\mathrm{B}(z),
\label{eq:omegaeom} \\
\partial_{\mu}\rho^{3\mu\nu}(z)+m^{2}_{\rho}\rho^{3\nu}(z)
&=& \sum_\mathrm{B} \Gamma_{\rho\mathrm{B}}
\bar{\psi}_{\mathrm{B}}(z) I_{3\mathrm{B}}
\gamma^{\nu}\psi_{\mathrm{B}}(z). \nonumber\\\label{eq:rhoeom}
\end{eqnarray}
Then, replace the meson fields with their expectation values, namely
\begin{eqnarray}
\sigma(z) &\rightarrow& \langle\sigma(z)\rangle=\sigma,
\label{sigammean} \\
\omega_{\mu}(z) &\rightarrow& \langle\omega_{\mu}(z)\rangle =
\omega_0, \label{omegamean} \\
\rho^{3}_{\mu}(z) &\rightarrow& \langle\rho^{3}_{\mu}(z)\rangle =
\rho^3_{0}.\label{rhomean}
\end{eqnarray}
Then the renormalized Lagrangian density is converted to
\begin{eqnarray}
\mathcal{L}^{\mathrm{MF}} &=&{\sum_{\mathrm{B}}} \bar{\psi}_\mathrm{B}(z)
\big(i\bar{A}_{0\mathrm{B}}\partial_t\gamma^0 +
i\bar{A}_{1\mathrm{B}} \bm{\partial}\cdot \bm{\gamma} -
\bar{A}_{2\mathrm{B}} m_{\mathrm{B}}^{} \nonumber \\
&& +g_{\sigma\mathrm{B}}\sigma-g_{\omega\mathrm{B}}\omega_0 \gamma^0
- \Gamma_{\rho\mathrm{B}} \rho^3_{0}I_{3\mathrm{B}}
\gamma^0\big)\psi_{\mathrm{B}}(z) \nonumber \\
&& -\frac{1}{2}m^{\ast2}_{\sigma}\sigma^{2} +
\frac{1}{2}m_{\omega}^{2} \omega^{2}_0 +
\frac{1}{2}m_{\rho}^{2}(\rho^3_{0})^2 \nonumber \\
&& +\sum_{l=e^{-},\mu^{-}}\overline{\psi}_{l}
\left(i\gamma^{\mu}\partial_{\mu}-m_{l}\right)\psi^{}_{l}.
\label{eq:RMFLhyperon}
\end{eqnarray}
The equation of motion of baryon fields, namely
Eq.~(\ref{eq:eomnucleon}), is now simplified to
\begin{eqnarray}
&&\Big[i\bar{A}_{0\mathrm{B}}\partial_t\gamma^0 +
i\bar{A}_{1\mathrm{B}}\bm{\partial}\cdot
\bm{\gamma}-g_{\omega\mathrm{B}}\omega_0\gamma^0 -
\Gamma_{\rho\mathrm{B}}\rho^3_{0}I_{3\mathrm{B}}\gamma^{0}
\nonumber \\
&&-\Sigma^\mathrm{R}_{0}\gamma^0 -(\bar{A}_{2\mathrm{B}}
m_{\mathrm{B}}^{}-g_{\sigma\mathrm{B}}\sigma)\Big]
\psi_\mathrm{B}(z)=0, \label{fermionRMFeom}
\end{eqnarray}
where $\Sigma^\mathrm{R}_{0}$ is the time component of the
rearrangement term. Accordingly,
Eqs.~(\ref{eq:sigmaeom})$-$(\ref{eq:rhoeom}) are modified,
\begin{eqnarray}
\sigma &=& \sum_\mathrm{B}\frac{g_{\sigma
\mathrm{B}}}{m^{\ast2}_{\sigma}}\langle\bar{\psi}_\mathrm{B}(z)
\psi_\mathrm{B}(z)\rangle = \sum_\mathrm{B}\frac{g_{\sigma
\mathrm{B}}}{m^{\ast2}_{\sigma}} n^{*}_\mathrm{s\mathrm{B}},
\label{sigmaRMFeom} \\
\omega_{0} &=& \sum_\mathrm{B}\frac{g_{\omega
\mathrm{B}}}{m^{2}_{\omega}}\langle \psi^{\dag}_\mathrm{B}(z)
\psi_\mathrm{B}(z)\rangle = \sum_\mathrm{B}\frac{g_{\omega
\mathrm{B}}}{m^{2}_{\omega}}n^{*}_\mathrm{B},
\label{omegaRMFeom} \\
\rho^{3}_{0} &=& \sum_\mathrm{B} \frac{\Gamma_{\rho
\mathrm{B}}}{m^{2}_{\rho}}\langle\psi^\dag_\mathrm{B}(z)
I_{3\mathrm{B}}\psi_\mathrm{B}(z)\rangle =
\sum_\mathrm{B}\frac{\Gamma_{\rho\mathrm{B}}}{m^{2}_{\rho}}
I_{3\mathrm{B}}n^{*}_{\mathrm{B}}. \label{rhoRMFeom}
\end{eqnarray}
Here, a raised asterisk is used to denote the inclusion of quantum
many-body effects.
The renormalized scalar density is
\begin{eqnarray}
n^{\ast}_\mathrm{s} = \sum_\mathrm{B}n^{\ast}_{\mathrm{sB}},
\end{eqnarray}
where
\begin{eqnarray}
n^{\ast}_{\mathrm{sB}} &=& 2\int^{k_{\mathrm{FB}}}_0
\frac{d^3\mathbf{k}}{(2\pi)^3}
\frac{m^{\ast}_{\mathrm{B}}/\bar{A}_{0
\mathrm{B}}}{E^{\ast}_{\mathrm{FB}}(\mathbf{k})}, \\
E^{\ast}_{\mathrm{FB}}(\mathbf{k}) &=& \sqrt{\frac{\bar{A}^{2}_{1
\mathrm{B}}}{\bar{A}^{2}_{0\mathrm{B}}}\mathbf{k}^{2}+
m^{\ast2}_{\mathrm{B}}}, \\
m^\ast_{\mathrm{B}} &=& \frac{\bar{A}_{2\mathrm{B}}m_{\mathrm{B}} -
g^{~}_{\sigma\mathrm{B}}\sigma}{\bar{A}_{0\mathrm{B}}}{.}
\end{eqnarray}

\begin{table*}[htbp]
\footnotesize 
\renewcommand{\arraystretch}{1.5} 
\caption{{Simulated model parameters for the
$\sigma\omega\rho2$ model and the nuclear quantities computed
at the saturation density within it, as developed in Ref.~\cite{Zhu24}.} The
experimental data for empirical nuclear quantities are:
$n^{~}_{\mathrm{B0}}=(0.16\pm0.01)~\mathrm{fm}^{-3}$,
$E_{\mathrm{b}}=(-16\pm1)~\mathrm{MeV}$,
$m^{\ast}_{\mathrm{N}}/m^{~}_{\mathrm{N}}=(0.56-0.75)$, $K=(240\pm
20)~\mathrm{MeV}$, and $E_{\mathrm{s}}=(28-34)~\mathrm{MeV}$
\cite{Zhu24}. {Because of the uncertainty in the value of \(L_s\)
\cite{Tagami22}, we select three values: $L_{\mathrm{s}} =
60$, $L_{\mathrm{s}} =
80$, and $L_{\mathrm{s}}= 87.56~\mathrm{MeV}$}
\cite{Zhu24}. } \centering
\begin{tabularx}{\textwidth}{l*{6}{>{\centering\arraybackslash}X}}
\toprule \hline\hline Model & $g_{\sigma\mathrm{N}}$ &
$g_{\omega\mathrm{N}}$ & $g_{\rho\mathrm{N}}$ &
$m^\ast_\sigma/m^{~}_\sigma$ & $\Lambda^{~}_{\mathrm{c}}$ & $a_\rho$
\\ \midrule \hline
$\sigma\omega\rho2L60$ & 17.5758 & 11.3153 & 6.5393 & 1.9355 & 1.6011 & 0.4097 \\
$\sigma\omega\rho2L80$ & 17.5758 & 11.3153 & 6.5393 & 1.9355 & 1.6011 & 0.1123 \\
$\sigma\omega\rho2L87.56$ & 17.5758 & 11.3153 & 6.5393 & 1.9355 & 1.6011 & 0.0000 \\
\addlinespace 
\midrule

& $n^{~}_{\mathrm{B}0}$(fm$^{-3}$) & $E_\mathrm{b}$(MeV) &
$m^{\ast}_\mathrm{N}$(MeV)
& $K$(MeV) & $E_{\mathrm{s}}$(MeV) & $L_{\mathrm{s}}$(MeV) \\
\midrule \hline $\sigma\omega\rho2L60$ & 0.1597 & -16.4255 & 0.6462
& 220.5497 & 28.9727 & 60.0000 \\
$\sigma\omega\rho2L80$ & 0.1597 & -16.4255 & 0.6462 & 220.5497 &
28.9727 & 80.0000 \\
$\sigma\omega\rho2L87.56$ & 0.1597 & -16.4255 & 0.6462 & 220.5497 &
28.9727 & 87.5612 \\
\addlinespace 
\bottomrule \hline\hline
\end{tabularx}
\label{TAB1hyperon}
\end{table*}

The expectation value of the energy-momentum tensor in the rest
frame of the matter is diagonal, namely
\begin{eqnarray}
\langle T^{\mu\nu}\rangle = \mathrm{diag}(\epsilon,P,P,P).
\end{eqnarray}
The energy density $\epsilon=\langle T^{00}\rangle$ is
\begin{eqnarray}
\epsilon&=&\sum_\mathrm{B}2\int^{k_\mathrm{FB}}_0
\frac{d^3\mathbf{k}}{(2\pi)^3}E^{\ast}_{\mathrm{FB}}(\mathbf{k})
+\frac{1}{2}m^{\ast2}_{\sigma}\Big(\sum_\mathrm{B}\frac{g_{\sigma
\mathrm{B}}}{m^{\ast2}_\sigma}n^{*}_{s\mathrm{B}}\Big)^2
\nonumber \\
&&+\frac{1}{2}m^{2}_{\omega}\Big(\sum_\mathrm{B}\frac{g_{\omega
\mathrm{B}}}{m^2_\omega}n^{*}_\mathrm{B}\Big)^2+\frac{1}{2}
m^{2}_{\rho}\Big(\sum_\mathrm{B}\frac{\Gamma_{\rho
\mathrm{B}}}{m^2_\rho} I_{3\mathrm{B}}n^{*}_{\mathrm{B}}
\Big)^2 \nonumber \\
&&+\sum_{l=e^{-},\mu^{-}}\frac{1}{\pi^2}\int^{k_{l}}_0
\mathbf{k}^2d|\mathbf{k}|\sqrt{\mathbf{k}^{2}_{l} +
m^{2}_{l}},\label{energyhyperon}
\end{eqnarray}
and the pressure $P=\frac{1}{3}\langle T^{ii}\rangle$ is
\begin{eqnarray}
P &=& \sum_\mathrm{B}\frac{2}{3}\int^{k_{\mathrm{FB}}}_0
\frac{d^3\mathbf{k}}{(2\pi)^3}\frac{\bar{A}^{2}_{1\mathrm{B}}
\mathbf{k}^{2}/\bar{A}^{2}_{0\mathrm{B}}}{E^{\ast}_{\mathrm{F
B}}(\mathbf{k})} - \frac{1}{2}m^{\ast2}_{\sigma} \Big(\sum_{\mathrm{B}}
\frac{g_{\sigma \mathrm{B}}}{m^{\ast2}_\sigma}n^{*}_{s\mathrm{B}}
\Big)^{2}
\nonumber \\
&& +\frac{1}{2}m^{2}_{\omega}\Big(\sum_\mathrm{B}\frac{g_{\omega
\mathrm{B}}}{m^2_{\omega}}n^{*}_\mathrm{B}\Big)^2 +
\frac{1}{2}m^{2}_{\rho}\Big(\sum_\mathrm{B}\frac{\Gamma_{\rho
\mathrm{B}}}{m^{2}_{\rho}}I_{3\mathrm{B}}n^{*}_{ \mathrm{B}}
\Big)^2\nonumber \\
&& +\sum_{l=e^{-},\mu^{-}}\frac{1}{3\pi^2}\int^{k_{l}}_{0}
d|\mathbf{k}|\frac{\mathbf{k}^{4}}{\sqrt{\mathbf{k}^{2} +
m^{2}_{l}}}+n^{\ast}_\mathrm{F}\Sigma^{\mathrm{R}}_{0}.
\label{presurehyperon}
\end{eqnarray}
The energy density and pressure are functions of baryon density
$n^{\ast}_\mathrm{F}$ through the Fermi momenta $k_\mathrm{FB}$ of
each species. {We have checked that the EOS satisfies the
thermodynamic relationship $P =
n^{*2}_\mathrm{F}\partial(\epsilon/n^{*}_\mathrm{F})/\partial
n^{*}_\mathrm{F}$.}

The NS cores contain, in addition to neutrons, a small fraction of
protons and electrons. As the density $n^{\ast}_{\mathrm{F}}$ is
sufficiently high, some electrons are replaced by muons when the
Fermi energy of electrons surpasses the rest energy of muons. In
that case, muons are energetically more favorable and the chemical
potentials satisfy $\mu_{e} = \mu_{\mu}$. As the density continues
to increase, hyperons could be excited in the inner core of NSs. The
formation threshold of the baryon $\mathrm{B}$ can be expressed
through the following relationship \cite{Glendenningbook}:
\begin{eqnarray}
\mu_{\mathrm{B}} &\geq& m^{\ast}_{\mathrm{B}}+g_{\omega
\mathrm{B}}\omega_{0} + g_{\rho \mathrm{B}}
I_{3\mathrm{B}}\rho_{0(3)} + \Sigma^\mathrm{R}_{0}.
\label{exciterelationship}
\end{eqnarray}
The baryon chemical potentials $\mu_{\mathrm{B}}$ are determined by
the conditions of $\beta$ equilibrium \cite{Glendenningbook}
\begin{eqnarray}
\mu_{n} &=& \mu_{\Lambda} = \mu_{\Xi^0} =
\mu_{\Sigma^0},\label{chemcon1} \\
\mu_{p} &=& \mu_{\Sigma^+} = \mu_{n}-\mu_{e}, \label{chemcon2} \\
\mu_{\Sigma^-}&=&\mu_{\Xi^-}=\mu_{n}+\mu_{e},\label{chemcon3}
\end{eqnarray}
where the chemical potentials of baryons, electrons, and muons are
given by
\begin{eqnarray}
\mu_{\mathrm{B}} &=&
\sqrt{\frac{\bar{A}^2_{1\mathrm{B}}}{\bar{A}^2_{0\mathrm{B}}}
k_{\mathrm{FB}}^{2} + m^{\ast2}_{\mathrm{B}}}+g_{\omega
\mathrm{B}}\Big(\sum_{\mathrm{B}}\frac{g_{\omega \mathrm{B}}}{m^{2}_{\omega}}
n^{\ast}_{\mathrm{B}}\Big) \label{chemicalpotential}\nonumber \\
&& +g_{\rho \mathrm{B}}I_{3\mathrm{B}}\Big(\sum_{\mathrm{B}}\frac{g_{\rho
B}}{m^{2}_{\rho}} I_{3\mathrm{B}}n^{\ast}_{\mathrm{B}}\Big) +
\Sigma^\mathrm{R}_{0}, \\
\mu_{e} &=& \sqrt{k^{2}_{e}+m^{2}_{e}}, \\
\mu_{\mu} &=& \sqrt{k^{2}_{\mu}+m^{2}_{\mu}}.
\end{eqnarray}
Here, $k_{e}$, and $k_{\mu}$ denote the Fermi momenta of electrons
and muons, respectively. Notice that the many-body effects are
incorporated in these chemical potentials. The lepton densities are
related to the corresponding Fermi momenta via the relation
$n_{e,\mu}=k^3_{e,\mu}/(3\pi^2)$. In addition, {the NS core should
obey the baryon number conservation and preserve electric charge
neutrality}, which are described by two identities,
\begin{eqnarray}
&&n^{\ast}_{\mathrm{F}} = \sum_{\mathrm{B}}
n^{\ast}_{\mathrm{B}} = \sum_{\mathrm{B}}\frac{k^{3}_\mathrm{FB}}{3
\pi^2\bar{A}_{0\mathrm{B}}},
\label{barcon} \\
&&\sum_{\mathrm{B}}Q_{\mathrm{B}}+\sum_{l}Q_{l} = \sum_{\mathrm{B}}\frac{q_{\mathrm{B}}
k^{3}_\mathrm{FB}}{3\pi^{2}\bar{A}_{0B}}-\sum_{l}
\frac{k^3_l}{3\pi^2} = 0. \label{neucon}
\end{eqnarray}
Here, $Q_{\mathrm{B}}$ and $Q_{l}$ represent the electric charges
carried by baryons and leptons, respectively, and {$q_{\mathrm{B}}$
is the electric charge of the baryon $\mathrm{B}$ in units of the
elementary charge}. Making use of
Eqs.~(\ref{exciterelationship})$-$(\ref{neucon}), we determine the
densities of baryons and leptons at a given total density
$n^{\ast}_{\mathrm{F}}$, which then generates a more realistic HS
EOS that satisfies the $\beta$ equilibrium, using
Eqs.~(\ref{energyhyperon}) and (\ref{presurehyperon}).

\begin{figure}[htbp]
\centering
\includegraphics[width=3.3in]{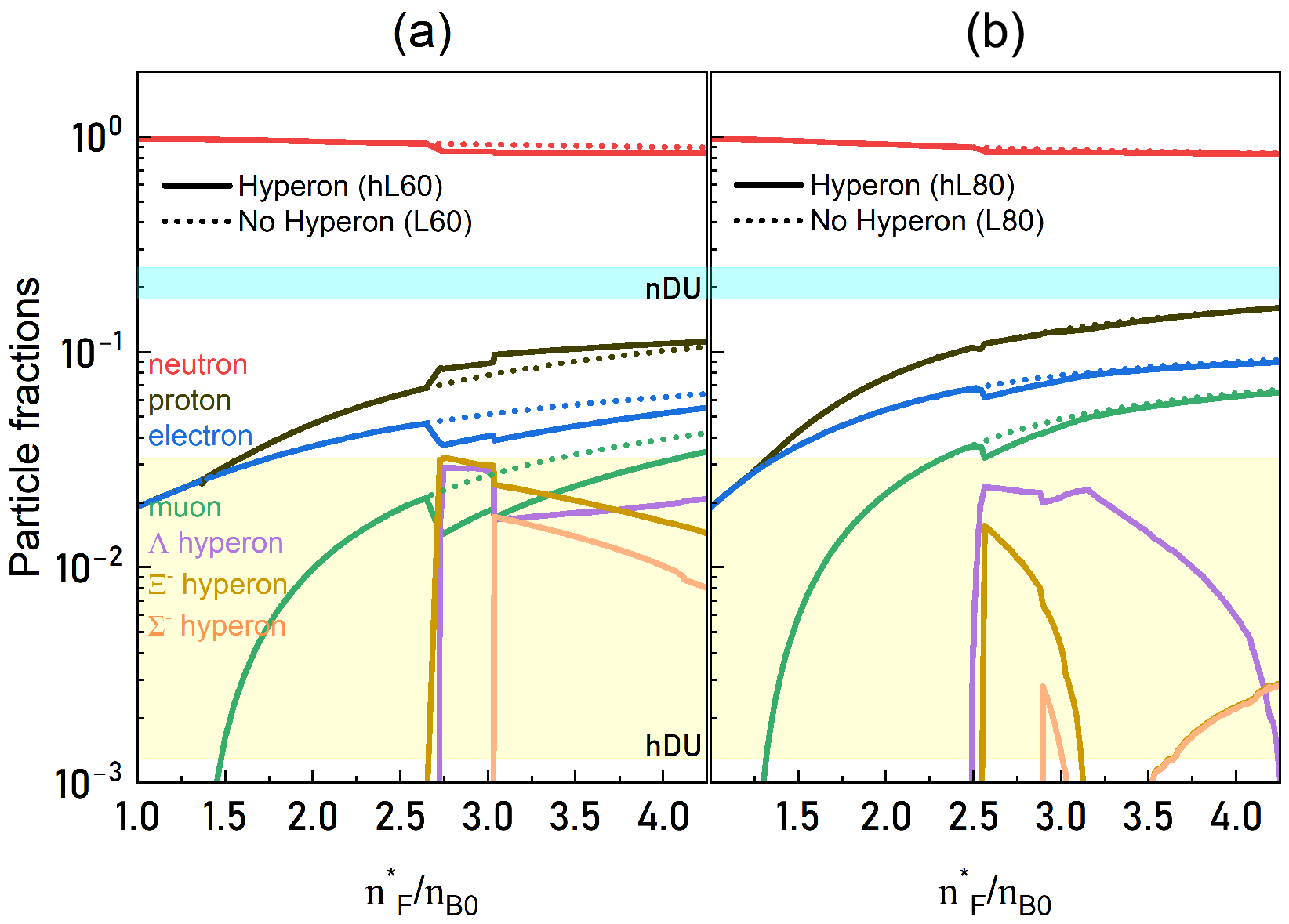}
\caption{{Particle fractions versus baryon density
$n^{\ast}_{\mathrm{F}}$ in units of $n_{\mathrm{B}0}$. (a) The hL60
(L60) model refers to the $\sigma\omega\rho2L60$ model with
(without) hyperons. (b) The hL80 (L80) model refers to
$\sigma\omega\rho2L80$ model with (without) hyperons. Shadowed
regions marked with nDU (hDU) illustrate the uncertain lower limit
of the threshold fractions for nucleonic (hyperonic) DU processes.}}
\label{fig:LParticlefrac}
\end{figure}

\section{Maximum mass $M_{\mathrm{max}}$ \label{sec:maximummass}}

We adjust the six parameters $g_{\sigma\mathrm{N}}$,
$g_{\omega\mathrm{N}}$, $g_{\rho\mathrm{N}}$,
$m^{\ast}_{\sigma}/m_{\sigma}$, $\Lambda_{\mathrm{c}}$, and
$a_{\rho}$ to fit six typical nuclear quantities, including the
nuclear saturation density $n^{}_{\mathrm{B}0}$, binding energy
$E^{}_\mathrm{b}$, effective nucleon mass $m^{{\ast}}_{\mathrm{N}}$,
{compressibility modulus} $K$, symmetry energy $E_{\mathrm{s}}$, and
symmetry energy slope $L_{\mathrm{s}}$ at the saturation density of
symmetric nuclear matter. Table~\ref{TAB1hyperon} shows that varying
$a_{\rho}$ does not affect the first five nuclear quantities, but
has a remarkable effect on $L_{\mathrm{s}}$. In Table
\ref{TAB1hyperon}, we list three sets of adjustable parameters,
referred to as the $\sigma\omega\rho2L60$ model with $L_s =
60~\mathrm{MeV}$, the $\sigma\omega\rho2L80$ model with $L_s =
80~\mathrm{MeV}$, and the $\sigma\omega\rho2L87.56$ model with $L_s
= 87.56~\mathrm{MeV}$.

We consider three hyperon species: $\Lambda$, $\Xi^{-}$, and
$\Sigma^{-}$. The $\Lambda$ and $\Xi^{-}$ are the first strange
baryons to appear with increasing baryon density. Although $\Sigma$
hyperons are generally disfavored by a repulsive potential
\cite{Drago14b}, we include the $\Sigma^{-}$ to evaluate its impact.
In Fig.~\ref{fig:LParticlefrac}, we present the dependence of
particle fractions on the normalized baryon density
$n^{\ast}_{\mathrm{F}}/n_{\mathrm{B}0}$ for four models: hL60, hL80,
L60, and L80. Obviously, models with different $L_{\mathrm{s}}$'s
exhibit distinct sequences of appearance for $\Lambda$ and
$\Xi^{-}$. For the hL60 model, shown in
Fig.~\ref{fig:LParticlefrac}(a), the fractions of $\Lambda$ and
$\Xi^{-}$ begin to noticeably emerge as the baryon density
increases, with $\Xi^{-}$ appearing at a critical density of
$2.70n_{\mathrm{B}0}$, followed by $\Lambda$ at
$2.75n_{\mathrm{B}0}$. This sequence differs from the hL80 model,
shown in Fig.~\ref{fig:LParticlefrac}(b), in which $\Lambda$ appears
first at $2.50n_{\mathrm{B}0}$, and $\Xi^{-}$ emerges shortly after
at $2.55n_{\mathrm{B}0}$. Meanwhile, the $\Sigma^{-}$ hyperon
appears only at much higher densities, around $3.00n_{\mathrm{B}0}$,
in both hL60 and hL80 models. Hence, the value of $L_s$ has an
obvious impact on the critical density for the emergence of
hyperons. Notably, even at densities above $4n_{\mathrm{B}0}$, the
fractions of $\Lambda$, $\Xi^{-}$, and $\Sigma^{-}$ all remain below
$0.02$, which justifies the consideration of only these three
hyperons.

\begin{figure}[htbp]
\includegraphics[width=3.86in]{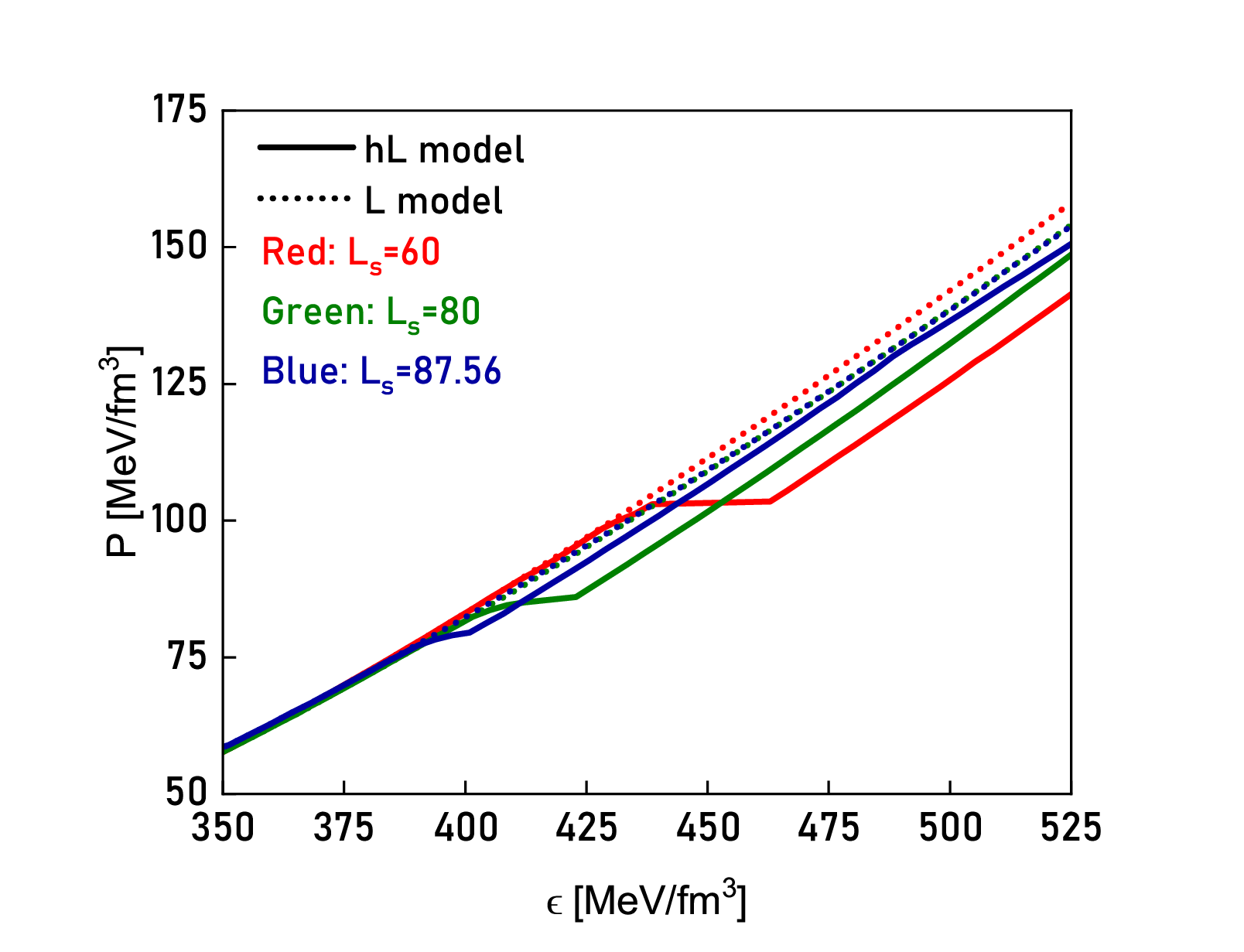}
\caption{Comparison between HS EOS and NS EOS in terms of the
pressure $P$ versus the energy density $\epsilon$.}
\label{fig:EOSCompare}
\end{figure}

A comparison between the $\epsilon$-$P$ curves obtained for HSs
(solid lines) and pure NSs (dashed lines) is depicted in
Fig.~\ref{fig:EOSCompare}. Clearly, the inclusion of hyperons
softens the EOS at high-energy densities. {Notice that the EOS,
particularly for models with smaller values of the symmetry energy
slope $L_{\mathrm{s}}$ (e.g., hL60), exhibit a dramatic abrupt
softening. This behavior is caused by the sudden and concurrent
emergence of $\Lambda$ and $\Xi^{-}$ hyperons within a narrow
density interval, as illustrated by Fig.~\ref{fig:LParticlefrac}(a).
The tuning parameter $a_{\rho}$, which governs the density
dependence of isovector meson coupling parameter
$\Gamma_{\rho\mathrm{B}}$, plays a crucial role. As shown in
Table~\ref{TAB1hyperon}, $L_{\mathrm{s}}$ is reduced as $a_{\rho}$
becomes larger, indicating that the symmetry energy increases more
slowly with growing density. This decreases the energy cost for
converting nucleons into hyperons at high densities and leads to a
sudden increase in the hyperon fractions. Consequently, once the
formation threshold (\ref{exciterelationship}) is crossed, multiple
hyperon species ($\Lambda$ and $\Xi^{-}$) can appear almost
simultaneously and their abundances rise sharply, which leads to the
observed abrupt softening in the $\epsilon$$-$$P$ relation.
Therefore, the parameter $a_{\rho}$ regulates both the hyperon
fractions and the stiffness of hyperonic matter.}

\begin{figure}[htbp]
\includegraphics[width=3.6in]{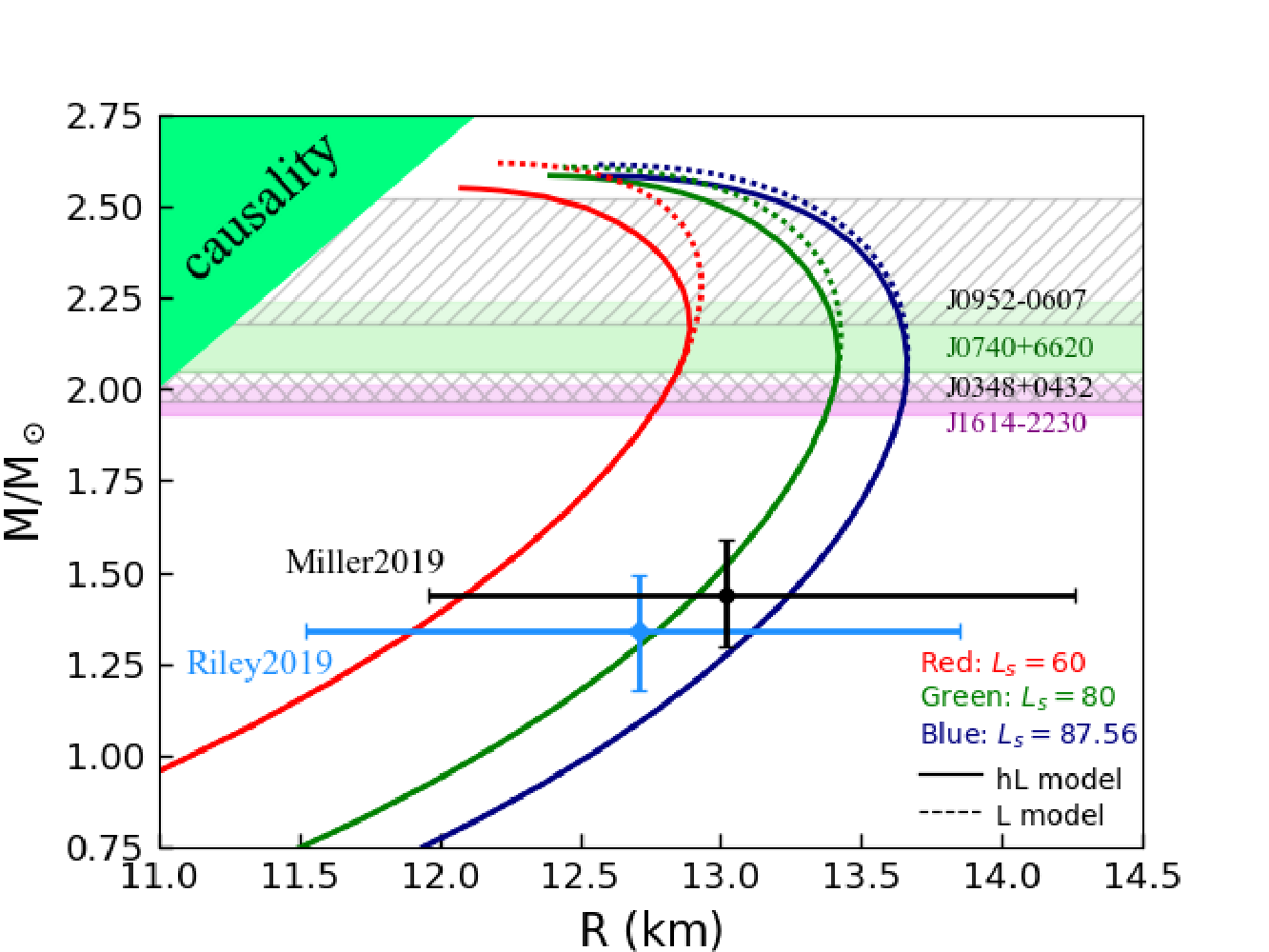}
\caption{{Comparison between the theoretical results of M-R
relations obtained from six models (hL60, hL80, hL87.56, L60, L80,
and L87.56) and some recent astrophysical observations of compact
stars.}} \label{fig:mr3}
\end{figure}

The mass-radius (M-R) relations are obtained by combining EOS with
the Tolman-Oppenheimer-Volkoff equations \cite{Tolman34,
Oppenheimer39}. To construct unified EOS, we use the \texttt{CUTER
v2} code \cite{Davis24, Davis25} to self-consistently reconstruct
the low-density crust EOS and match them to the high-density core
EOS, which ensures thermodynamic consistency and smoothness across
the crust–core transition. In Fig.~\ref{fig:mr3}, we compare the
M-R curves calculated based on six different models with the
empirical data extracted from astrophysical observations of compact
stars. Noticeably, the radii for low-mass NSs inferred from our
M–R curves are in full agreement with NICER constraints on PSR
J0030+0451 \cite{Miller19, Riley19}. The M-R relations predict
smaller radii for low-mass NSs when $L_{s}$ takes smaller values.
This trend holds true irrespective of whether the hyperons are
included, since the properties of low-mass NSs are nearly unaffected
by hyperons that emerge at densities around
$2n_{\mathrm{B}0}$-$3n_{\mathrm{B}0}$ in the NS core.

In a number of existing solutions to the hyperon puzzle based on
RMFT \cite{Colucci13, Sun23, Providencia19, Tu25, Fortin20,
Fortin16}, the maximum mass $M_{\mathrm{max}}$ has been elevated to
the range of $2.2 M_{\odot} - 2.3 M_{\odot}$ through careful
selections of appropriate hyperon-meson interactions. While such a
mass range suffices to explain most high-mass NSs, PSR J0952-0607
remains an exception. {In a recent work \cite{Kumar23}, the high
mass of PSR J0952-0607 was explained within some pure nucleonic RMFT
models. These models invoke a set of extended nonlinear self- and
cross-coupling terms \cite{Kumar23} to stiffen the EOS at high
densities, leading to maximum NS masses in the range $\sim
(2.34$-$2.50) M_{\odot}$. As in other RMFT studies, the nonlinear
terms are handled within the static Hartree approximation. These
static terms furnish a classical background potential for the
baryons as their quantum fluctuations are neglected. When a baryon
is scattered by this potential, its energy is not changed. This
implies that the baryons are nearly independent: each feels the same
potential yet none affects another. The static potential shifts the
baryon masses and chemical potentials, but otherwise the baryons
propagate freely, forming an almost noninteracting degenerate
fermion gas rather than a truly correlated quantum many-body system.
After hyperons are included, the resulting EOS will be drastically
softened even with the same nonlinear meson couplings
\cite{Dhiman07}, which decreases the maximum HS mass well below the
$\sim (2.34$-$2.50) M_{\odot}$ range.}

Compared to those RMFT studies, our approach treats the baryon-meson
interactions nonperturbatively and retains their explicit time
(i.e., energy) dependence. In our scheme, all baryons are mutually
correlated via the exchange of dynamical, fluctuating mesons.
Quantum many-body effects enter the EOS through the averaged
quantities $\bar{A}_{0,1,2}$, whereas the overall effect of
nonlinear meson couplings is encoded in a single renormalized
$\sigma$ mass. The resulting HS EOS are stiff enough to support a
maximum mass of $M_{\mathrm{max}} \approx 2.59M_{\odot}$, even
without considering stellar rotation or strong magnetic fields. This
value is more than adequate to account for the masses of several
known massive NSs, including PSR J1614-2230, PSR J0348+0432, and PSR
J0740+6620. Remarkably, it also supports the exceptionally high mass
of PSR J0952-0607, {which is challenging for other hyperonic
scenarios lacking many-body effects.}

Our stiff EOS still respects causality after including hyperons.
Explicitly, we have verified that the sound of speed $c_{s}$ is
smaller than the speed of light $c$ across all densities relevant to
the stable stellar sequences in Fig.~\ref{fig:mr3}, irrespective of
whether hyperons are present or not.

\section{Fate of fast cooling \label{sec:cooling}}

In addition to the insufficient value of $M_{\mathrm{max}}$, RMFT
results face another issue about the fast cooling of HSs
\cite{Maslov15}. The proton fraction obtained from RMFT calculations
typically exceeds $0.15$, larger than the threshold fraction
$Y_{\mathrm{nDU}}$ for the nucleonic DU process \cite{Lattimer91}.
Meanwhile, the hyperon fractions are within the range of 0.20$-$0.60
\cite{Providencia19, Li18, Drago14, Drago14b, Sedrakian23,
Weissenborn12, Wei24, Fortin20, Miyatsu13, Lopes14, Zhang18, Tu25},
much higher than the threshold fraction $Y_{\mathrm{hDU}}$ for the
hyperonic DU processes \cite{Prakash92}. A direct implication of
these results is that HSs, in their early stages, would likely be
significantly colder than what is inferred from astrophysical
observations. This situation remains unchanged when the regulating
role of baryon pairing is taken into account \cite{Yakovlev01,
Page04}. As shown in \cite{Potekhin20}, the observed age-temperature
relations for dozens of NSs with low masses can be well explained by
the standard cooling framework that precludes all DU processes.
There are also evidences \cite{Negreiros18, Raduta18} suggesting
that the observed cooling data of many NSs can only be understood if
DU processes occur in NSs with masses $> 1.8M_{\odot}$. In a few
RMFT models, such as NL$3\omega\rho$ \cite{Fortin20, Fortin21}, DU
processes are avoided in low-mass HSs, but they inevitably occur in
HSs having high and intermediate masses. Moreover, the corresponding
$M_{\mathrm{max}}$ \cite{Fortin20, Fortin21} is not large enough to
account for PSR J0952-0607.

In our scenario, the proton and hyperon fractions are substantially
suppressed due to the quantum many-body effects, which is in stark
contrast to RMFT results. {In the absence of hyperons, a comparison
between L60 and L80 models reveals that the proton fraction
increases with density at a rate proportional to $L_{\mathrm{s}}$.
For $L_{\mathrm{s}}=80~\mathrm{MeV}$, although the proton fraction
increases toward the threshold range of $0.11$–$0.15$
\cite{Lattimer91} at higher densities, it remains below this
critical interval, and is therefore insufficient to trigger the
nucleonic DU process. In the case of
$L_{\mathrm{s}}=60~\mathrm{MeV}$, however, the reduced
$L_{\mathrm{s}}$ together with quantum many-body effects results in
a much slower rise of the proton fraction with density. After
introducing the hyperons, the proton fraction in the $L_{\mathrm{s}}
= 80~\mathrm{MeV}$ model is almost unchanged, whereas in the
$L_{\mathrm{s}} = 60~\mathrm{MeV}$ model it is noticeably enhanced.
Nevertheless, it still remains smaller than $0.11$ for all baryon
densities below $4.5 n_{\mathrm{B}0}$, as shown in
Figs.~\ref{fig:LParticlefrac}(a) and~\ref{fig:LParticlefrac}(b).
Thus, the nucleonic DU processes are entirely inhibited. Initially,
the hyperon fractions rise after their onset and reach a maximum,
but they subsequently decline with further increasing density.
Eventually they stabilize below $0.02$, a trend clearly shown for
both values of $L_{s}$.} The threshold fraction for the hyperonic DU
processes is extremely low and highly uncertain, estimated to be
within the range of $0.0013$-$0.0320$ \cite{Prakash92}. Comparing
these two values reveals that the hyperonic DU processes may or may
not occur in HSs. Although these processes cannot be entirely ruled
out in our scenario, their likelihood of occurrence has been reduced
to an unprecedented low level. As a result, fast cooling normally
does not occur in HSs, even for those with high masses.

\begin{figure}[htbp]
\includegraphics[width=3.66in]{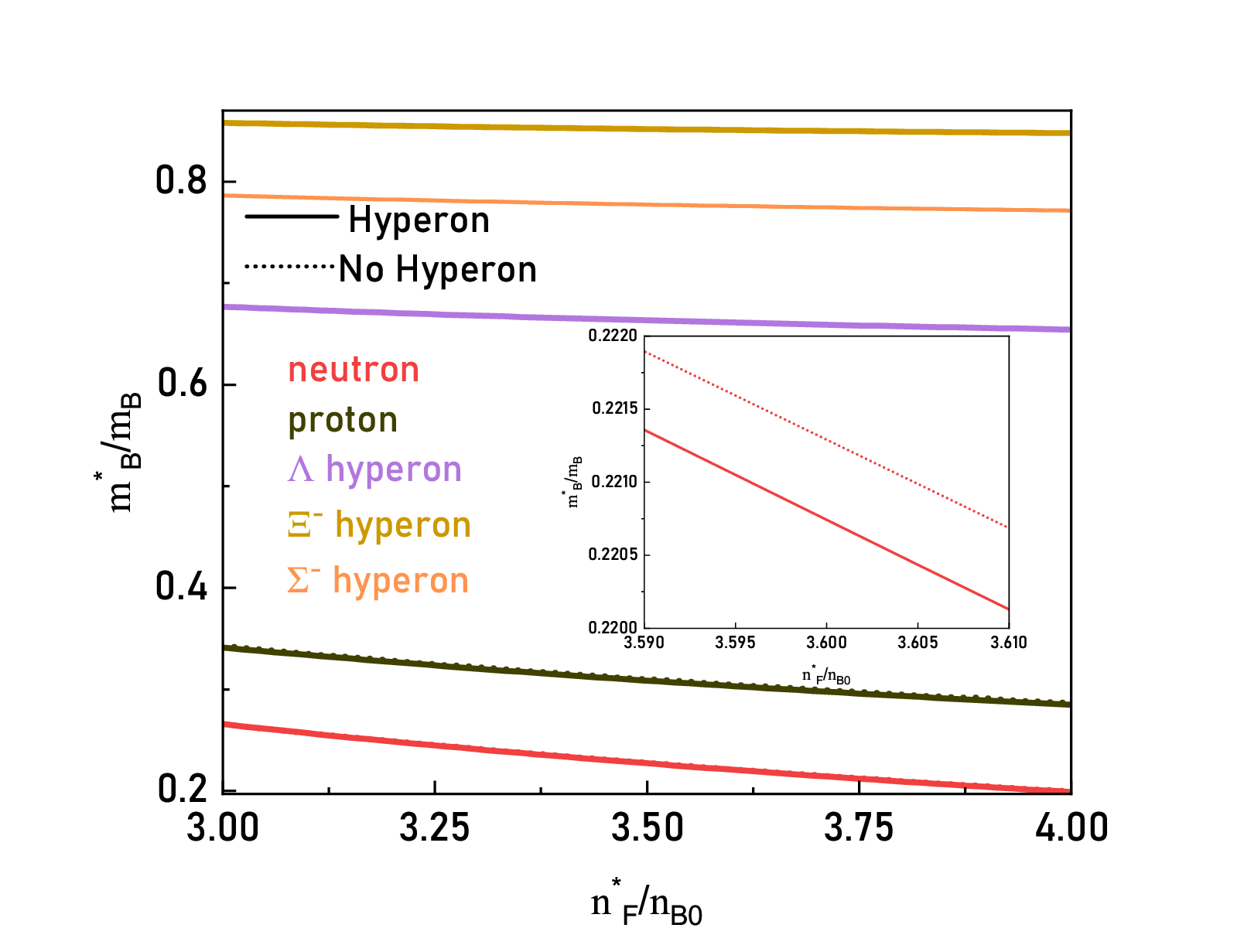}
\caption{Ratios of effective baryon masses to bare masses versus
normalized baryon density $n^{\ast}_{\mathrm{F}}/n_{\mathrm{B}0}$
for the $\sigma\omega\rho L80$ model.} \label{fig:MassCompare}
\end{figure}

To understand why the neutron fraction is much higher than proton and hyperon fractions, it is useful to examine the density dependence of effective baryon masses. As shown in Fig.~\ref{fig:MassCompare}, in the absence of hyperons, the many-body effects already reduce the effective neutron mass $m^{\ast}_n$ to small values at high densities. Once hyperons are included, $m^{\ast}_n$ is further decreased, dropping from $0.265m^{\ast}_n$ at $3.00n_{\mathrm{B}0}$ to $0.200m^{\ast}_n$ at $4.00n_{\mathrm{B}0}$. Thus, the neutrons become increasingly relativistic at higher densities. The renormalized baryon density of each species can be rephrased as
\begin{eqnarray}
n^{\ast}_{\mathrm{B}} &=& \frac{k^{3}_\mathrm{FB}}{3
\pi^2\bar{A}_{0\mathrm{B}}} = \frac{\left(\mu^{\ast2}_{\mathrm{B}} -
m^{\ast2}_{\mathrm{B}}\right)^{\frac{3}{2}}}{3\pi^2
\bar{A}_{1\mathrm{B}}}.\label{effemass}
\end{eqnarray}
The effective baryon chemical potential obtained from
Eq.~(\ref{chemicalpotential}) is given by
\begin{eqnarray}
\mu^{\ast}_{\mathrm{B}} &=& \mu_{\mathrm{B}}-g_{\omega
\mathrm{B}}\Big(\sum_{\mathrm{B}}\frac{g_{\omega \mathrm{B}}}{m^{2}_{\omega}}
n^{\ast}_{\mathrm{B}}\Big) \nonumber \\
&& -g_{\rho \mathrm{B}}I_{3\mathrm{B}}\Big(\sum_{\mathrm{B}}\frac{g_{\rho
B}}{m^{2}_{\rho}} I_{3\mathrm{B}}n^{\ast}_{\mathrm{B}}\Big) -
\Sigma^\mathrm{R}_{0} \nonumber\\
&=&\sqrt{\frac{\bar{A}^2_{1\mathrm{B}}}{\bar{A}^2_{0\mathrm{B}}}
k_{\mathrm{FB}}^{2} + m^{\ast2}_{\mathrm{B}}}.
\label{effechemical}
\end{eqnarray}
Based on Eq.~(\ref{effemass}), one can infer that a reduction in
$m^{\ast}_n$ enhances the neutron density, which is essential to
maintain a high neutron fraction. Protons display a similar trend
but with a considerably higher effective mass, leading to a much
lower proton fraction. The effective hyperon masses are lowered as
baryon density rises, but remain above $0.65$ times bare masses.
Thus, hyperons are nonrelativistic, which suppresses their
fractions.

According to the above analysis, our current work presents a
distinct prediction concerning the cooling rate of HSs, particularly
those with high masses, compared to that of RMFT studies. It appears
that measuring the age-temperature relations of massive HSs offers
the most efficient means to verify which prediction is more
reliable. The currently available observational data of such
relations are rather limited and insufficient to draw a conclusive
answer. It is hoped that the age-temperature relation of NSs with
masses $> 2.0M_{\odot}$ could be obtained with an acceptable
precision in the near future. This would help determine whether the
presence of hyperons in NSs necessarily leads to fast cooling.

To make the above qualitative prediction more testable, we will
carry out quantitative cooling simulations in the future. These
calculations would produce two sets of cooling curves for massive
HSs$-$one without and one with quantum many-body effects. A direct
comparison of the two sets of cooling curves will then allow us to
quantify how many-body correlations influence the thermal evolution
of HSs.

It is worth mentioning that a few exceptional NSs are observed to
exhibit rapid cooling. A prominent example is the young isolated NS
in the Cassiopeia A supernova remnant \cite{Ho09, Ho10, Posselt18,
Posselt22}. The rapid cooling of this particular NS can be
attributed to the enhanced neutrino emission resulting from Cooper
pairing breaking and formation \cite{Page11, Shternin11} in
conjunction with the associated superfluid and superconducting
quantum critical phenomena \cite{Zhu2410}, without the need to
invoke DU processes.

\section{Summary and discussion \label{sec:summary}}

Our results demonstrate that incorporating the quantum many-body
effects into the EOS of HSs not only yields a maximum mass
sufficient to account for the masses of all the observed NSs, but
also prevents the fast cooling induced by DU processes. Thus, our
scenario provides a unified solution to {the two facets of the
hyperon puzzle}.

While our calculated {$M_{\mathrm{max}} \approx 2.59M_{\odot}$} is
already quite large, it may be further increased if the effects of
NS spin \cite{Yuan05, Silva25} and strong magnetic field
\cite{Broderick02} are considered. Moreover, strong magnetic field
can change the particle fractions within HSs \cite{Broderick02,
Yue09}, which would alter the conditions of DU processes. These
issues will be addressed in future works. It is also interesting to
apply the DS equation framework to examine the impact of isobars on
the M-R relation \cite{Drago14, Drago14b, Li18, Sedrakian23,
Silva25} and other quantities.

In the present work, the DS integral equations of the
renormalization functions $A_{0,1,2}(\varepsilon)$ are solved using
the bare-vertex approximation to baryon-meson couplings. With this
approximation, the coupling parameters $g_{\sigma \mathrm{B}}$,
$g_{\omega \mathrm{B}}$, and $g_{\rho \mathrm{B}}$ are taken as
constants. The good agreement between our results and available
astrophysical observations indicates that this truncation scheme
captures the essential physics of dense matter, yet upgrading to
energy-momentum-dependent baryon-meson vertices would yield a better
description of the EOS and lead to improved results for the maximum
mass and thermal evolution. Retaining the full energy-momentum
dependence substantially increases the computational cost. Thus, a
significantly more efficient algorithm must be developed to solve
the more complicated DS equations.

\section{Acknowledgment}

We thank the anonymous referees for constructive suggestions that
helped improve the manuscript. H.F.Z. thanks Wei Liu, Li Ma, Jinzhi
Shen, and Zhipeng Zhang for valuable discussions on numerical
calculations. G.Z.L. thanks Ang Li and Zhonghao Tu for helpful
discussions. H.F.Z. and X.W. are supported by the National Natural
Science Foundation of China (Grants No.~12433002 and No.~12073026).
H.F.Z. is also supported by the China Postdoctoral Science
Foundation (Grant No.~2025M783436). Y.F.Y. is supported by the
National Natural Science Foundation of China (Grants No.~12433008
and No.~12393812) and the National SKA Program of China (Grant
No.~2020SKA0120300). H.F.Z, X.W., and Y.F.Y. acknowledge the support
by the Cyrus Chun Ying Tang Foundations, the 111 Project for
Observational and Theoretical Research on Dark Matter and Dark
Energy (B23042), and Professor Yipeng Jing's Academician
Workstation. The numerical calculations in this paper have been done
on the supercomputing system in the Supercomputing Center of
University of Science and Technology of China.

\end{document}